\documentclass[aps,prb,twocolumn,superscriptaddress,preprintnumbers,amsmath,amssymb,floatfix,showkeys]{revtex4}

\usepackage[dvips]{graphicx}
\usepackage{xspace}
\usepackage[usenames,dvipsnames]{xcolor}
\usepackage{array}
\usepackage{ulem}
\usepackage{hyperref}
\usepackage[english]{babel}
\usepackage{lipsum}
\usepackage{calc}

\begin{document}

\title{Dynamical resonance quench and Fano interference in spontaneous Raman scattering from quasiparticle and collective excitations}

\author{J.~Zhu}
\author{R.~B.~Versteeg}%
\author{P.~Padmanabhan}%
\author{P.~H.~M.~van Loosdrecht}%
\email[Corresponding author:]{pvl@ph2.uni-koeln.de}

\affiliation{II. Physikalisches Institut, Universit\"{a}t zu K\"{o}ln, Z\"{u}lpicher Stra{\ss}e 77, D-50937 K\"{o}ln, Germany}

\date{\today}

\begin{abstract}
Time-resolved spontaneous Raman spectroscopy serves as a probe for incoherent quasiparticle and collective excitation dynamics, and allows to distinguish symmetry changes across a photoinduced phase transition through the inelastic light scattering selection rules. Largely unexplored is the role of the Raman resonance enhancement in the time-domain, and the transient interaction between scattering from quasiparticles and collective excitations, with the latter interaction leading to a Fano interference.

In this work, we report on the observation of a phonon Raman tensor quench and Fano interference after strong photoexcitation of an intrinsic semiconductor. We observed a dynamic phonon scattering rate asymmetry and spectral asymmetry through simultaneous detection of both the anti-Stokes and Stokes response. The asymmetric phonon scattering rate is ascribed to the combined effect of the transient phonon population and the reduction of the phonon Raman tensor resulting from the photoexcited hole population. This same hole population results in a strong enhancement of the Fano spectral asymmetry. 
\end{abstract}



\maketitle

A quantum-mechanical description of the nature and transport properties of the various quasiparticles and collective excitations in semiconductors is integral to characterizing the physical properties of a wide array of technologically relevant materials.\cite{bhattacharya2011comprehensive,hase2003birth} Hot carrier and phonon relaxation dynamics in semiconductors, \cite{hase2010interaction,othonos1998} for example, plays a central role in our understanding of carrier transport and energy dissipation in electronic and optoelectronic devices. As such, an accurate interpretation of the quasiparticle and collective excitation interactions and dynamics is vital not only from a fundamental perspective, but also to enable new applications.

Over the past four decades, a range of ultrafast spectroscopic probes have provided us insight into carrier and phonon relaxation dynamics and their respective time-scales in semiconductors. \cite{othonos1998,sabbah2000measurement} These measurements often rely on the coherent excitation of phonons, of which the dynamics is consecutively probed in the modulation of the refractive index. \cite{sabbah2002prb,riffe2007prb,kato2011apl98} A more direct view of phonon dynamics is provided by time-resolved inelastic scattering techniques. \cite{trigo2013,hannah2013,faustibook,versteeg2018} This has the advantage over coherent excitation techniques that one can follow the true time evolution of a photoexcited material through its incoherent response, rather than the time evolution of coherent excitations.

Time-resolved spontaneous Raman scattering is a reasonably well-established probe for incohorent phonon population dynamics in semiconductors, \cite{kash1985,oberli1987,tsen1986,tsen1988,letcher2007,kttsen2006,faustibook} and vibrational dynamics in molecules. \cite{malinovsky2012slow,iwata2007local,sahoo2011time,matsuda2003time,petterson2015tissue,hamaguchi1994,kruglik2011,faustibook} More recently the technique has been applied to study phonon dynamics in a range of carbon allotropes. \cite{kang2008optical,song2008,yan2009,kang2010,chatzakis2011temperature,yang2017novel,zhu2018,faustibook} Through detailed balance of the anti-Stokes I$_{\rm AS}$ to Stokes I$_{\rm S}$ scattering intensity ratio the (phonon) population number $n$ can be directly quantified by I$_{\rm AS}$/I$_{\rm S}$\,$\propto$\,$n/(n+1)$. This relation stems from the fluctuation dissipation theorem, \cite{loudon1978time} which on ultrafast timescales ($\ll$\,ps) may lose its validity due to the presence of truly non-thermal occupation statistics. The inelastic light scattering selection rules $\vert \hat{\rm E}_{\rm S} \cdot\chi^{''}_A\cdot \hat{\rm E}_{\rm I} \vert ^2$ for incoming electric field $\hat{\rm E}_{\rm I}$ and scattered field $\hat{\rm E}_{\rm S}$ entail the symmetry $A$ of the Raman-scattering channel $\chi ^{''}_A$. \cite{compaan1984} This may be used to probe symmetry changes across a photoinduced phase transition, as was demonstrated for a high-to-low structural symmetry transition in antimony, \cite{fausti2009} and melting of the superconducting condensate in a high-temperature cuprate superconductor.\cite{saichu2009} Two largely unexplored aspects to time-resolved spontaneous Raman are the role of the Raman resonance enhancement in the time-domain, \cite{zhu2018} and the dynamic interaction between quasiparticle and collective excitation scattering, with the latter interaction leading to a Fano interference.

Here we demonstrate a photo-induced resonance quench, and Fano-interference in the transient phononic and electronic spontaneous Raman response of an intrinsic indirect band gap semiconductor. An intrinsic (100) oriented silicon sample (resistivity \textgreater$10000$\,$\Omega\cdot$\,cm) was used for the experiments. Time-resolved Raman scattering experiments were performed using a $740$\,nm optical pump pulse with a temporal width of $0.3$\,ps. The transient Raman spectra are recorded in a backscattering geometry using a $512$\,nm, $1.5$\,ps probe pulse. \cite{versteeg2018} The polarization of both the pump and probe beams were aligned with the [110] crystallographic direction of the sample and the cross polarized scattered light was measured. The steady-state Raman spectrum of silicon is well-known, the main features being the optical phonon peak centered at $\pm$\,$520$\,cm$^{-1}$, and, in $p$-doped silicon, a broad and relatively weak electronic continuum extending to several hundred cm$^{-1}$. This feature is attributed to inter-valence band Raman scattering between the uppermost heavy- and light-hole valence bands.\cite{cerdeira1973ssc,cerdeira1973prb,chandrasekhar1978,burke2010raman} Coupling of the optical phonon to this continuum leads to a pronounced Fano asymmetry \cite{fano2005,fano1961} of the optical phonon Raman response. These coupled spectral features lead to an intricate time-domain evolution following optical excitation, \cite{letcher2007,kato2018} obscuring the pure population dynamics. As we will see, however, the detection of both the Stokes and anti-Stokes responses allows for the determination of the transient optical phonon population dynamics, while simultaneously yielding information on the temporal-evolution of the Raman scattering efficiency. Interestingly, the optically induced valence band hole density is found to lead to a transient enhancement of the Fano effect which can be observed in real time. 

\begin{figure}[t]
\center
\includegraphics[scale=.48]{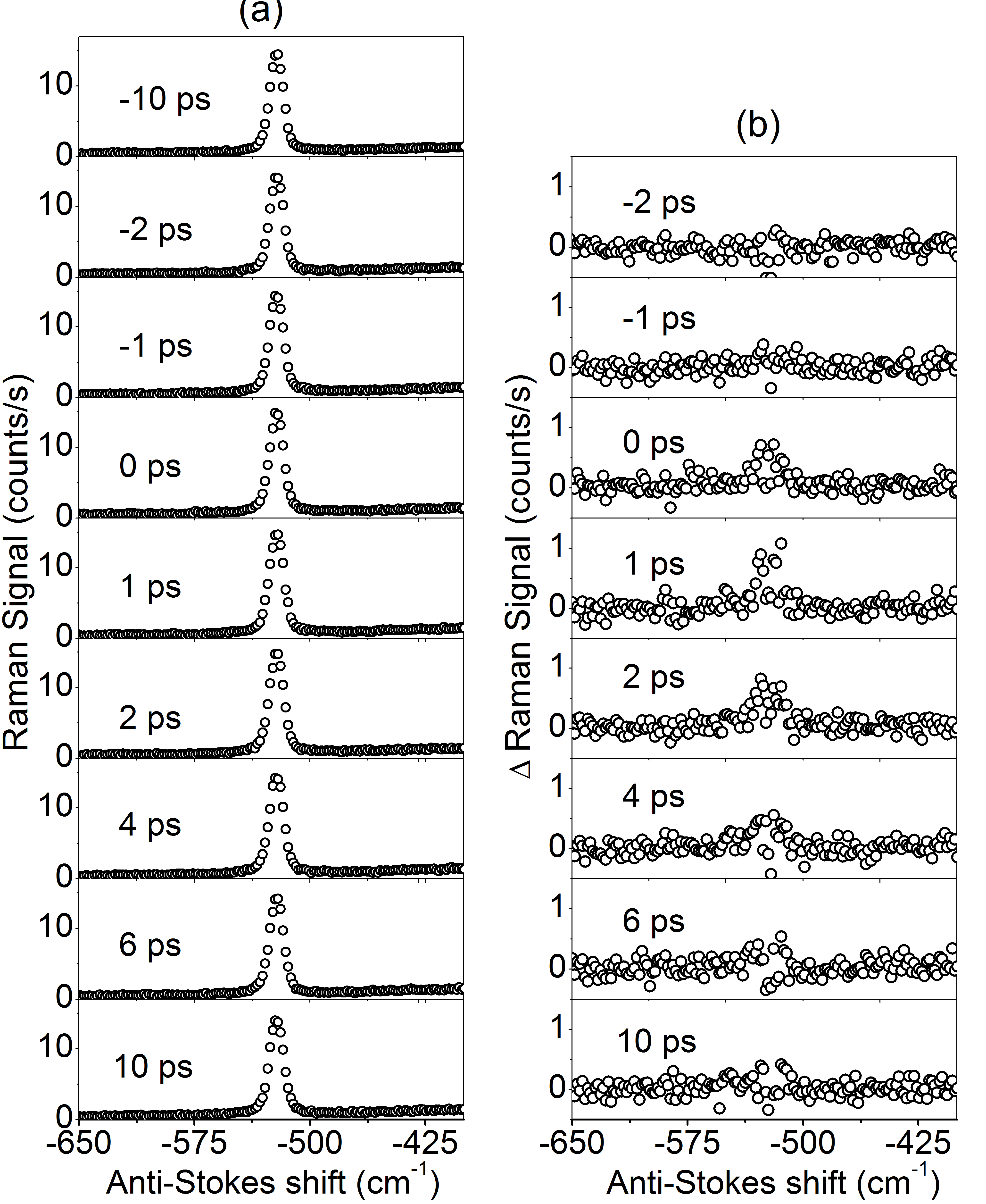}
\caption{Time-resolved spontaneous silicon anti-Stokes Raman spectra recorded at a carrier density of $\sim$\,$1.8\times 10^{18}$\,cm$^{-3}$. (a) Raman spectra at various delay times before and after optical excitation. (b) Difference spectra obtained by subtracting the spectra at $-10$\,ps from the spectra in (a).}
\label{fig:trantistokes}
\end{figure}

Figure \ref{fig:trantistokes} shows the transient anti-Stokes Raman spectra of silicon with a photoexcited carrier density of around $\sim$\,$1.8\times 10^{18}$\,cm$^{-3}$ measured at various pump-probe delay times.\cite{footnote} For this relatively low transient carrier density, one observes a fast increase of the intensity near the optical phonon response at $-520$\,cm$^{-1}$ which subsequently decreases with time until it vanishes after $6$\,-\,$10$\,ps. This corresponds to the expected creation of an optical phonon population through the electron-phonon interaction, which subsequently decays on a ps timescale through anharmonic coupling with acoustic phonons.\cite{klemens1966,kang2010,kang2008optical,song2008,chatzakis2011temperature}

\begin{figure}[t]
\center
\includegraphics[scale=.48]{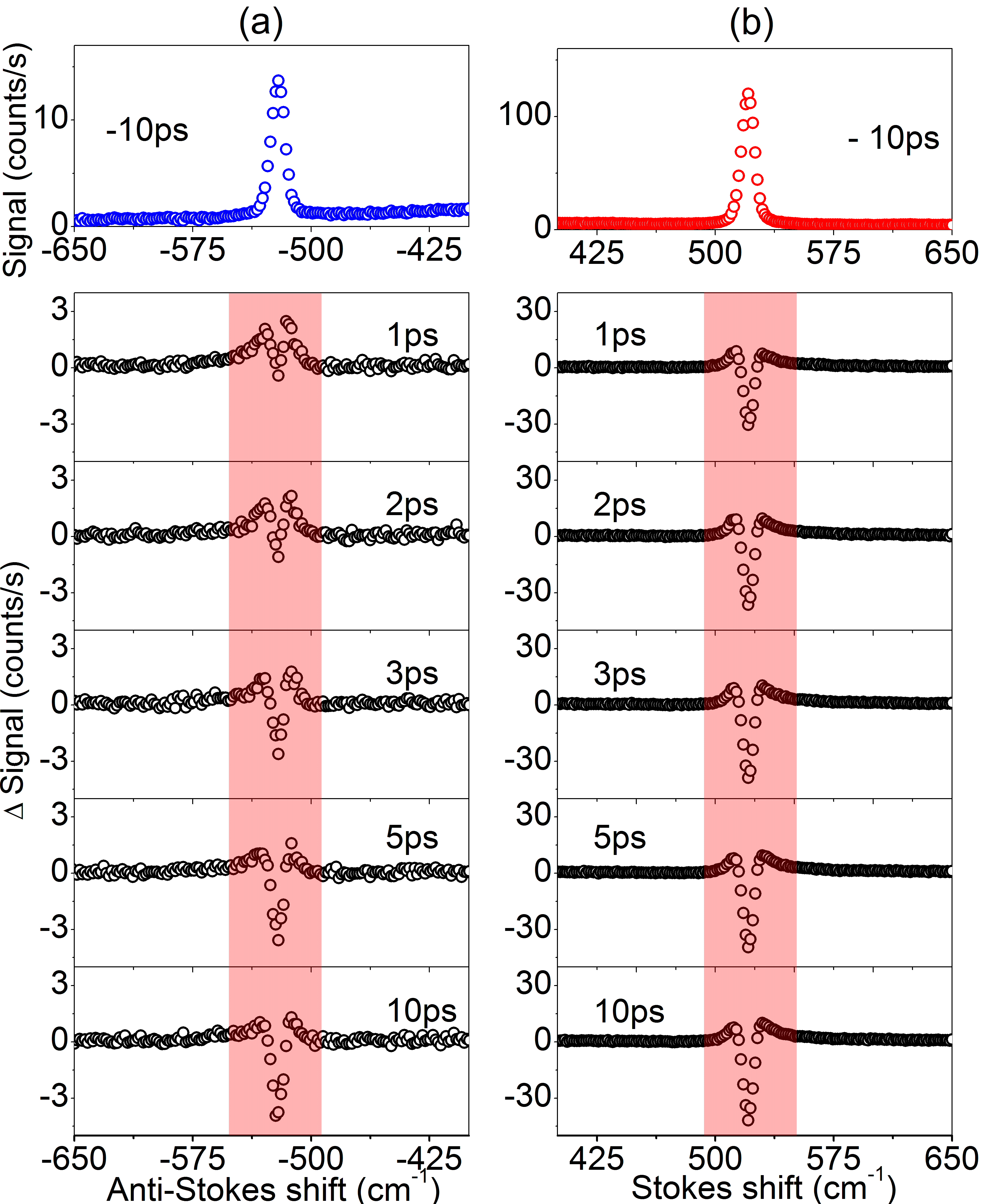}
\caption{Lower panels: Time-resolved spontaneous silicon Raman difference spectra recorded at a carrier density of $\sim$\,$1.8\times 10^{19}$\,cm$^{-3}$ for anti-Stokes (a) and Stokes (b) scattering. These spectra are obtained by subtracting the spectra recorded for $-10$\,ps delay (top panels) from the spectra recorded at positive pump-probe delays.}
\label{fig:diffspectra}
\end{figure}

However, this simple picture is incomplete, as demonstrated in Fig.\,\ref{fig:diffspectra} at higher excited carrier density ($\sim$\,$1.8\times 10^{19}$\,cm$^{-3}$) from both anti-Stokes (left panels) and Stokes (right panel) difference Raman spectra. Particularly striking is the non-trivial substructure and the notable asymmetry between the Stokes and anti-Stokes spectra.  At early delay times ($1$\,-\,$2$\,ps) the anti-Stokes difference spectra show a positive response near the phonon frequency, with slight negative difference signal in the middle, while a negative difference signal dominates on the Stokes side. For longer time delays, the negative signal on both sides gains strength until the Stokes and anti-Stokes signals become symmetric with a negative component in the middle and positive shoulders on the sides at $\sim$\,$10$\,ps. The negative phonon difference signal persists for several nanoseconds, which is on the order of the carrier population lifetime. These spectral features originate from mixed effects of the optically induced phonon and hole populations, spectral broadening, and a Fano interference with the hole continuum. \cite{kato2018} We first focus on the observed changes in the overall signal intensity. 

\begin{figure}[t]
\center
\includegraphics[scale=.48]{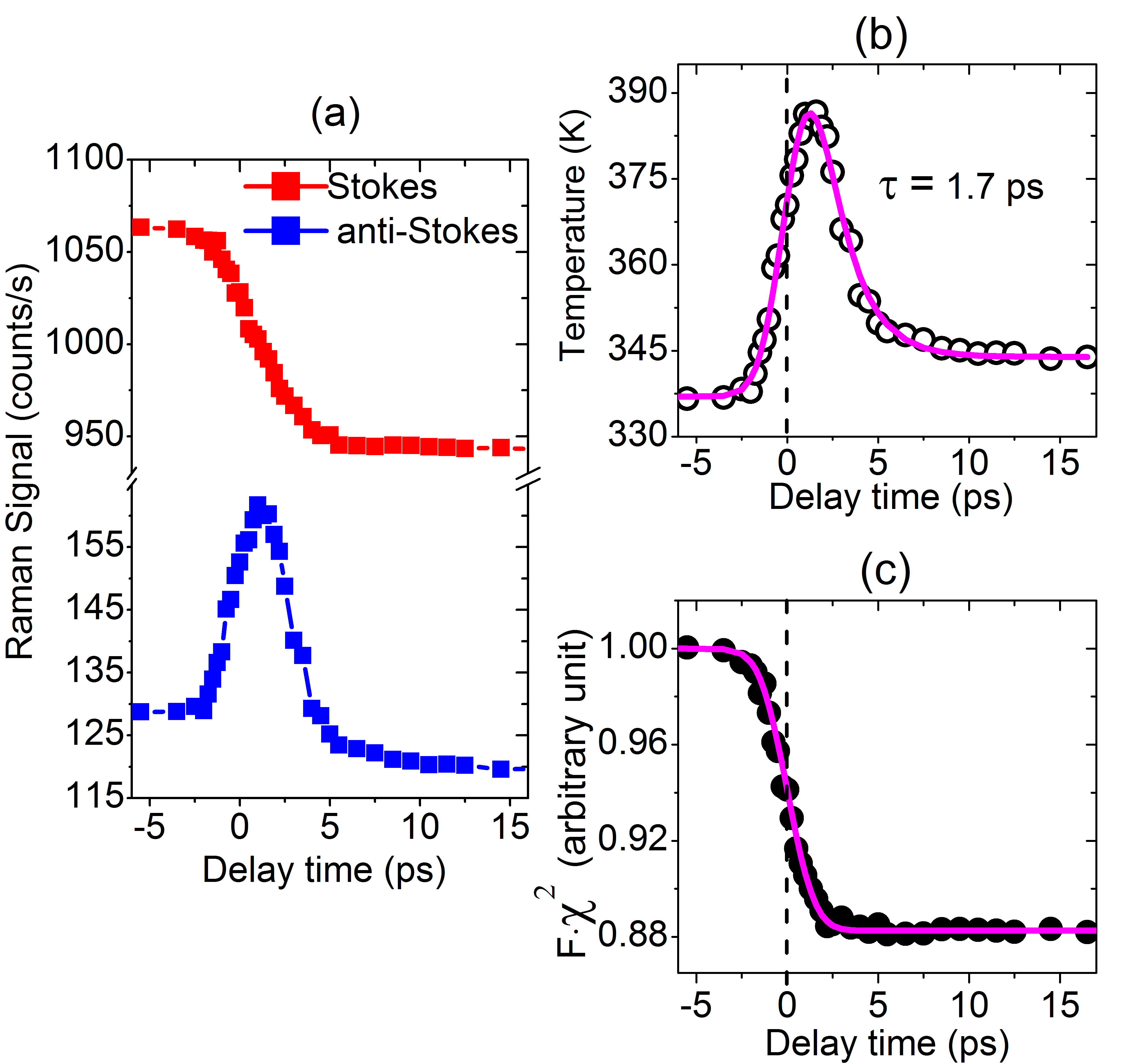}
\caption{Relaxation dynamics of phonon equivalent temperature and Raman scattering strength at a carrier density of $\sim$\,$1.8\times 10^{19}$\,cm$^{-3}$. (a) Decay of the optical phonon intensity for Stokes scattering (red squares) and anti-Stokes scattering (blue squares). (b) Extracted phonon temperature evolution. (c) Time-evolution of the Raman scattering strength $F\cdot\chi^2$.}
\label{fig:transients}
\end{figure}

The spectra from Fig.\,2 were integrated over the optical phonon contribution ($490$\,cm$^{-1}$ to $550$\,cm$^{-1}$) and corrected by subtracting the increasing electronic background. \cite{tanaka1993} The resulting relaxation dynamics are shown in Fig.\,\ref{fig:transients}a. The Stokes signal shows an immediate decrease in the scattering intensity after optical excitation. In contrast, the anti-Stokes signal first shows a rapid increase of scattering intensity which decreases only at later time delays. To understand this behavior, we note that the Raman signal for a specific phonon mode is determined by the population $n_p$, the Raman tensor $\chi^2$, and the related optical coefficient function $F$. It can be expressed as

\begin{equation}
{\rm I}_{\rm S}=C\cdot F_S\cdot\omega_S^3\cdot\chi^2_S\cdot(1+n_p)
\end{equation}
\noindent and 
\begin{equation}
{\rm I}_{\rm AS}=C\cdot F_{AS}\cdot\omega_{AS}^3\cdot\chi^2_{AS}\cdot n_p 
\end{equation}

\noindent for the Stokes and anti-Stokes response, respectively. \cite{compaan1984} The factor $C$ includes the factors which are equal for both sides, such as the excitation laser energy dependence and the excitation being scattered from ({\it i.e.} the phonon frequency). The optical coefficient function $F$ can be written as

\begin{equation}
F_S=\frac{\mathcal{T}_S}{(\alpha_L+\alpha_S)\cdot \eta_S}
\end{equation}

\noindent and

\begin{equation}
F_{AS}=\frac{\mathcal{T}_{AS}}{(\alpha_L+\alpha_{AS})\cdot \eta_{AS}}
\end{equation}

\noindent for Stokes and anti-Stokes scattering, respectively. Here $\mathcal{T}$, $\alpha$ and $\eta$ are the transmission coefficient, absorption coefficient and refractive index at the corresponding photon energies. These optical constants and the Raman tensor are strongly resonance dependent, and usually depend on temperature, which complicates the determination of phonon population based on the above formula. However, as demonstrated in Ref.\,\onlinecite{compaan1984}, these correction factors are close to unity well below the resonance energy ($3.45$\,eV). As such, the following relation holds:

\begin{equation}
F_{\rm S}\cdot\chi^2_{\rm S}\,\approx\,F_{\rm AS}\cdot\chi^2_{\rm AS}\,\equiv\,F\cdot\chi^2
\end{equation}

\noindent Using this relation and formula (1) and (2), one can now determine the phonon population and the phonon temperature from the experimental data. The resulting phonon temperature and the scattering strength $F\cdot\chi^2$ are plotted in Fig.\,\ref{fig:transients}b and \ref{fig:transients}c. Before time zero, the phonon temperature is $\sim$\,$340$\,K, showing an average heating above room temperature. Upon excitation, the phonon temperature rapidly increases to $\sim$\,$390$\,K after which it decays close to the average value within a few picoseconds. An exponential fit gives a lifetime of approximately $1.7$\,ps (solid line in Fig.\,\ref{fig:transients}b), consistent with the expected decay dynamics of optical phonons into acoustic phonons. \cite{klemens1966,kash1985,letcher2007,kang2010}

More surprising is the rapid decrease of about $10$\% in the scattering strength $F\cdot\chi^2$ upon excitation. A convolution of a step function and a Gaussian with half width of our setup resolution fits well (solid line in Fig.\,\ref{fig:transients}c) to the curve. Apparently, the reduction of the scattering strength can either originate from changes in the optical coefficient function $F$ (which is dictated by the optical constants), or from the Raman tensor square $\chi^2$ immediately after photoexcitation. Generally, the change of optical constants upon excitation are small. To confirm this, we performed a reflectivity measurement with identical excitation conditions, which shows that the changes of reflectivity upon excitation is less than $0.2$\,\%.\cite{footnote2} This small change in reflectivity, indicating a negligible change of the refractive index and thus the other derived optical constants, is not sufficient to explain the observed large change in scattering strength $F\cdot\chi^2$.

\begin{figure}[t]
\center
\includegraphics[scale=.48]{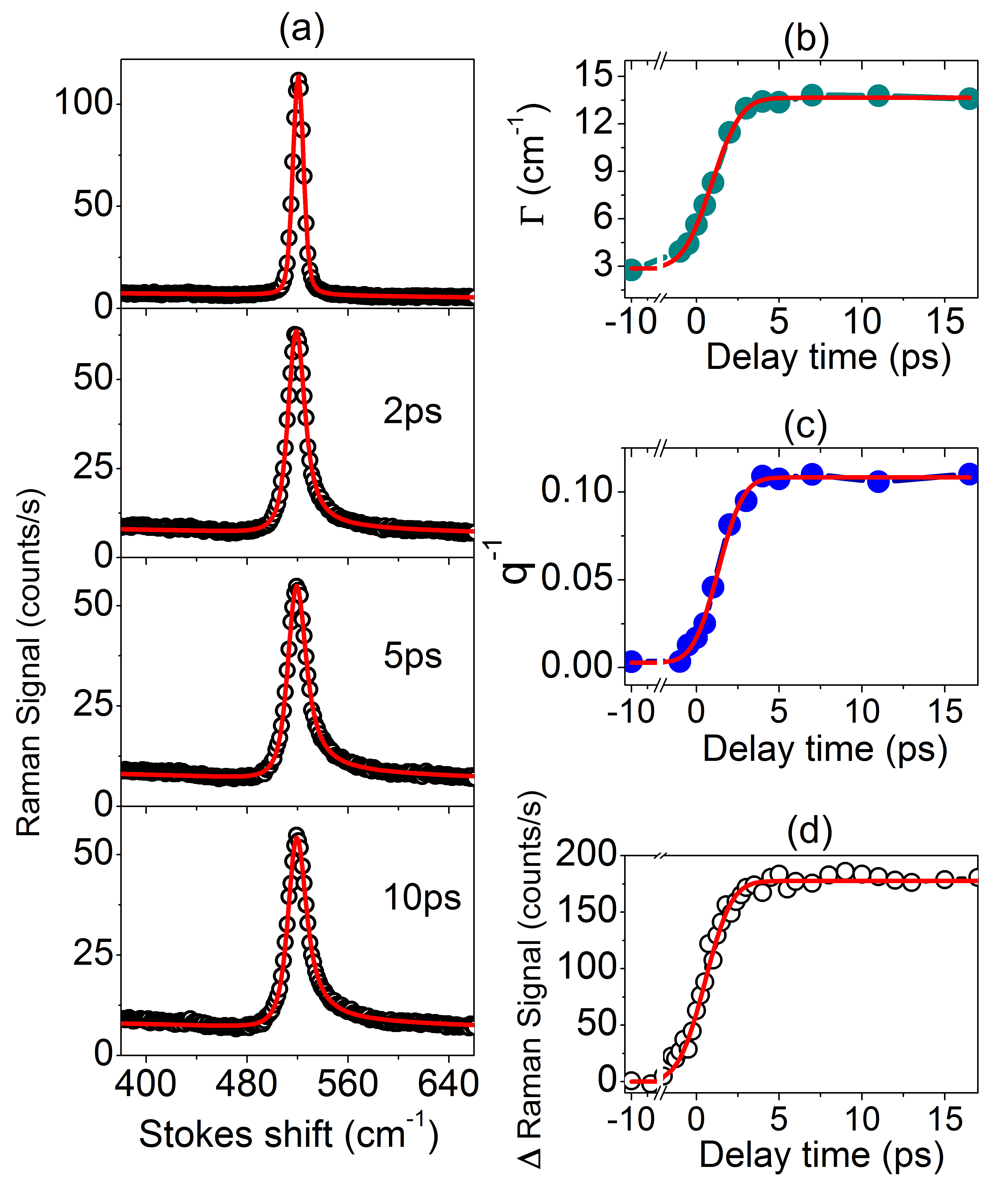}
\caption{Transient Fano interference and hole population dynamics at a photoinduced density of $\sim$\,$4.3\times 10^{19}$\,cm$^{-3}$. (a) Time dependent Raman spectra (circles) and Fano-fits (red lines).  (b) Fitted line-width parameter $\Gamma$. (c) Fitted inverse Fano parameter $q^{-1}$. (d) Hole scattering dynamics integrated in the spectral range from $650$\,cm$^{-1}$ to $750$\,cm$^{-1}$.}
\label{fig:fano}
\end{figure}

A more likely scenario for the observed decrease in scattering strength $F\cdot\chi^2$ arises from the resonant nature of the Raman scattering in silicon at the Raman probe excitation energies used in this study. The scattering is resonant to vertical transitions from the topmost valence band to the lowest conduction band in the $\Gamma$\,$-$\,$L$ direction in the Brillouin zone, and is thereby proportional to charge population. 
\cite{renucci1975} In addition to this, excitonic effects are thought to play a role as well.\cite{gillet2013} 
The pump laser pulse optically induces a hole density in the valence band, as evidenced by the observed changes in the hole continuum scattering (Fig.\,\ref{fig:fano}d). \cite{tanaka1993} The optically induced charge density could lead to changes in the excitonic effects on the Raman resonance discussed in Ref.\,\citenum{gillet2013}, which would lead to a strongly probe pulse energy dependent change in $\chi^2$. However, our experiments show that the optically induced changes in the Raman spectra do not depend on probe energy within our experimental accuracy. \cite{footnote3} Excitonic effects therefore do not appear to play a large role in the tensor quench. Obviously the resonant enhancement also strongly depends on the number of available initial and final states. Here, in particular the hole population in the vicinity of the $\Gamma$-point is of interest. \cite{renucci1975} Recent quantum theoretical work on silicon has shown that a few percent of the valence band states will be unoccupied at a photoexcitation density of $\sim$\,$10^{19}$\,cm$^{-3}$, (Ref.\,\onlinecite{sangalli2016}) which would lead to a reduction of $\chi^2$ of about $10$\,\% (Refs.\,\citenum{footnote5} and \citenum{boyd2008}). This is in good agreement with our observations, and we hence attribute the transient reduction of the Raman resonance to the photoinduced hole density.

We now turn to the line shape of the phonon response, and hole dynamics. Fano interference between the optical phonon and the electronic hole continuum in $p$-doped silicon is expected to be weak for the $512$\,nm probe wavelength.\cite{cerdeira1973ssc,cerdeira1973prb,burke2010raman} However, at higher excitation densities, the changes in the Fano interference induced by the optical excitation are observable, \cite{kato2018} as can be seen in Fig.\,\ref{fig:fano}a, which shows the transient Raman Stokes spectra (circles) at an excitation density of $\sim$\,$4.3\times 10^{19}$\,cm$^{-3}$ for various time delays. Generally the Fano line-shape can be described by

\begin{equation}
{\rm I}\propto\frac{(q+\epsilon)^2}{1+\epsilon^2}
\end{equation}

\noindent where $q$ is the so-called Fano parameter which is inversely proportional to the spectral density of the electronic continuum, and $\epsilon$\,$=$\,$2(\omega-\omega_0)/\Gamma$ with $\omega_0$ the bare phonon frequency, and $\Gamma$ the line width parameter, which apart from the bare phonon part, has a contribution due to the Fano interference proportional to the spectral density of the electronic continuum. \cite{cardonabook} The transient behavior of the inverse Fano parameter $q^{-1}$ and the line-width $\Gamma$ obtained from fits (solid lines in Fig. \,\ref{fig:fano}a) are depicted in Fig.\,\ref{fig:fano}b and \ref{fig:fano}c. Before time zero, the Raman peak resembles a Lorentzian line shape ($q^{-1}$\,$\approx$\,$0.003$),
whereas after photoexcitation the line-shape shows a pronounced Fano asymmetry ($q^{-1}$\,$\approx$\,$0.11$).
As expected, a nearly stepwise time scale of the initial response of both $\Gamma$ and $q^{-1}$ were observed, which are identical to the electronic continuum dynamics (Fig.\,\ref{fig:fano}d). Since these parameters are all governed by the dynamics of the optically induced long-lived holes, no decay of these induced effects is observed within the measured time window. Finally, we note the values of the Fano parameters in our pump-probe experiments of pure silicon are consistent with results obtained from steady-state Raman scattering silicon for similar hole-doping levels. \cite{footnote4,fano2005,fano1961,cerdeira1973ssc,cerdeira1973prb,chandrasekhar1978,burke2010raman}

In summary, time-resolved spontaneous Raman scattering experiments on an intrinsic semiconductor show a prominent dynamic asymmetry between both the intensity and line-shape of the phononic Stokes and anti-Stokes responses. These asymmetries originate from the interplay between the transient phonon population dynamics, a quench of the phonon Raman resonance due to a photoinduced hole population, and an enhanced Fano interference due to the same hole population, in line with simple expectations. The present results provide a deeper understanding of the spontaneous Raman-scattering response of semiconductors to strong optical excitation, and expand the use of of time-resolved spontaneous Raman scattering in the study of collective excitation and quasiparticle dynamics in solid state materials.

This project was partially financed by the Deutsche Forschungsgemeinschaft (DFG) through SFB-1238 project B05, and instrument grant INST217/782-1. We gratefully acknowledge fruitful discussions with Daniele Fausti. R.~B.~V. acknowledges partial funding through the Bonn-Cologne Graduate School of Physics and Astronomy (BCGS). 

J.~Z. and R.~B.~V. contributed equally to this work.

\end{document}